\documentclass{emulateapj}

\usepackage{apjfonts}
\usepackage{amsmath}
\usepackage{epstopdf}

\shortauthors{Constantin et al. }  

\shorttitle{The M 94 conundrum}

\begin{document}

\slugcomment{to appear in Advances in Astronomy} 

 
 
 \title{M94 As A Unique Testbed for Black Hole Mass Estimates and AGN Activity At Low Luminosities\altaffilmark{1,2}} 

\altaffiltext{1}{Based on observations made with the NASA/ESA {\it
Hubble Space Telescope}, obtained from the data archive at the Space
Telescope Science Institute. STScI is operated by the Association of
Universities for Research in Astronomy, Inc., under NASA contract NAS
5-26555.}
\altaffiltext{2}{Some observations reported here were obtained at the MMT Observatory, a joint facility of the Smithsonian Institution and the University of Arizona.}

\author{Anca Constantin\altaffilmark{3} and Anil C. Seth\altaffilmark{4}}

\altaffiltext{3}{ James Madison University, Harrisonburg, VA 22807}
\altaffiltext{4}{University of Utah, Salt Lake City 84112}

\begin{abstract}

We discuss the peculiar nature of the nucleus of M94 (NGC 4736) in the context of new measurements of the broad H$\alpha$ emission from $HST$-STIS observations.  We show that this component is unambiguously associated with the high-resolution X-ray, radio, and variable UV sources detected at the optical nucleus of this galaxy.  These multi-wavelength observations suggest that NGC 4736 is one of the least luminous broad-line (type 1) LINERs, with $L_{\rm bol} = 2.5 \times 10^{40}$~erg s$^-1$.  This LINER galaxy has also possibly the least luminous broad line region known ($L_{H \alpha} = 2.2 \times 10^{37}$~erg s$^{-1}$).  We compare black hole mass estimates of this system to the recently measured $\sim 7\times 10^{6}~M_\sun$ dynamical black hole mass measurement.  The fundamental plane and $M - \sigma^*$ relationship roughly agree with the measured black hole mass, while other accretion based estimates (the $M-FWHM(H\alpha)$ relation, empirical correlation of BH mass with high-ionization mid IR emission lines, and the X-ray excess variance) provide much lower estimates ($\sim 10^5 M_\sun$).  An energy budget test shows that the AGN in this system may be deficient in ionizing radiation relative to the observed emission-line activity.  This deficiency may result from source variability or the superposition of multiple sources including supernovae.

\end{abstract}


\section{Introduction: LINERs and M94}

Most high mass galaxies are known to host massive black holes (BH),
some passively lurking in their centers while others are actively
accreting surrounding material (e.g., G\"{u}ltekin et al. 2009a and references therein).  The mechanism that causes a BH's
activity to turn on and off is still largely unknown.
Understanding the structure of the active galactic nuclei (AGN) at
their lowest luminosities is crucial to determining the physical and
possibly evolutionary links between the most luminous galaxy centers
and the passive ones.  However, at low luminosities it is difficult
to disentangle the various emission mechanisms that could be
concurrently present in galaxy centers.  As a consequence, the
dominant power source of a large majority of actively line-emitting
galaxies remains ambiguous (e.g., Ho 2008 for a review).

Diagnostic diagrams \citep{bpt, vei87, kau03, kew06} are relatively
successful in separating out bona-fide accretion sources (Seyferts)
from nuclei whose emission-line activity is mainly powered by young,
hot stars (H {\sc ii} galaxies), based on emission line ratios.  At
least 50\% of the strong line emitters fall easily onto the H {\sc ii}
locus, however, only less than 10\% are of the Seyfert type \citep{ho97V,kau03,con06}.  
A large fraction of the objects ``in between'' these two categories,
that exhibit relatively low levels of ionization (i.e., low values of
[\ion{O}{3}]/H$\beta$), maintain reasonably strong forbidden line
activity (i.e., high values of [\ion{N}{2}]/H$\alpha$), and are
classified as low ionization nuclear emission regions (LINERs).  The
reminder are usually called Transition objects (Ts).  Whether Ts and
Ls are powered, at least partly, by accreting BHs, and thus could be
called AGN, is a matter of continuous debate \citep{ho08}.

There are some typical emission characteristics that are considered to
be particularly good indications that accretion onto a massive BH is
an important, if not the dominant source of ionization in some Ls, and possibly Ts.
The detection of broad H$\alpha$ emission, regardless of its
strength or luminosity, is generally considered to be $the$ clue to
AGN emission.  Some LINERs (and maybe Ts as well) exhibit these
features, however, the majority of these systems show only narrow
emission, which could be generated by shocks, post-starbursts, or
other processes unrelated with accretion.  Observations outside of the
optical wavelengths often reveal AGN signatures in ambiguous and even
starburst galaxy nuclei.  X-rays are particularly good tracers of
accretion, however, they are not efficient in distinguishing AGN at
$L_X \la 10^{42}$ erg s$^{-1}$, where contamination by X-ray binaries
can be significant.  
X-rays are also unlikely to detect heavily absorbed AGN (i.e.,
Compton-thick; $N_{\rm H} > 1.5 \times 10^{24}$ cm$^{-2}$).  On the
other hand, mid-IR high ionization emission lines like [\ion{Ne}{5}]
$\lambda$14.32 $\mu$m, 24.32$\mu$m (97.1 eV) appear to be a
trustworthy indicator of AGN activity \citep{weedman05, armus06,
  abel08} due to the extreme conditions (i.e., very hard ionizing
radiation) required to produce them; because these features have
considerably lower optical depth, their detection can also reveal
Compton thick AGN \citep[e.g., NGC 1068]{sturm02}.  This technique
has now been applied to reveal new and large numbers of optically
unidentified AGN \citep{satyapal08, goulding09}, providing thus
sensitive improvements on previous AGN censuses.  These studies, along
with X-ray and radio studies of nearby galaxies
\citep[e.g.][]{nagar05,zhang09,desroches09} suggest that a majority of
LINERs and a large fraction of transition galaxies might in fact host
accreting black holes.
The presence of an accreting black hole does not however guarantee that the accretion power is the dominant source of ionization of those galaxy nuclei.  
A more recent assessment of the energy budget of LINERs by \citet{eracleous10}
argues that in 85\% of LINERs the AGN ionizing photons are not sufficient for producing the observed nebular emission, and thus other power sources are likely to dominate.

Some new potentially powerful insights into the excitation mechanism
of the low luminosity AGN (LLAGN), and in particular the ambiguous sources, come from recent
studies of large statistical samples of nearby galaxy nuclei, which
reveal a potential {\it H {\sc ii} $\rightarrow$ S $\rightarrow$ T
  $\rightarrow$ LINER $\rightarrow$ Passive Galaxies} evolutionary
sequence in the process of BH growth within galaxies \citep{con08,
  con09}. This sequence traces trends in (1) increasing host halo
mass, (2) increasing environmental density, (3) increasing central BH
mass and host stellar mass, (4) decreasing BH accretion rate, (5)
aging of the stellar population associated with their nuclei, and (6)
decreasing in the amount of dust obscuration, which might translate
into a decrease in the amount of material available for star-forming
or accretion.  In this picture, Seyferts and Ts are transition phases
between the initial onset of accretion, usually swamped by the
star-forming gas and associated dust, which is seen optically as an H
{\sc ii} system, and the final phase of accretion observed as LINERs
of already massive BHs.  While this idea is supported by various other
independent observational studies of low luminosity AGN and starburst
galaxies \citep{sch07, sch10}, along with state of the art hydrodynamical
models for the life-cycles of the most luminous AGN (e.g., Di Matteo
et al. 2005, Hopkins et al. 2006), it is very probable that not $all$
sources fit into this scenario.    
It is also important that this evolutionary sequence idea is tested on samples that span a narrow distribution in Hubble types;  because most of the H {\sc ii}s and Ts are relatively late-type disk galaxies that, likely, never experienced a (recent) major merger, it is possible that the trigger of such a sequence is different from the merger that initiates a similar life-cycle at high luminosities.  
Nevertheless, investigating the challenges certain objects bring to this idea is
useful for identifying and quantifying the caveats associated with
this sequence.  These objects may also be at interesting stages in
their galaxy evolution.


NGC 4736 (or M94, UGC 7996) is a captivating example of an ambiguous
galaxy nucleus, which poses challenges to the general understanding of
AGN phenomena, including the above mentioned sequence.  This object is
one of the closest ($d = 4.3$ Mpc; Tully \& Fisher 1988) nearly face-on spiral, with a
SAab Hubble Type.  Its proximity enables study of details that
would be unobservable in more distant systems.  Its nucleus has a
low-luminosity LINER spectrum, but has been also included in catalogs
of Transition objects \citep{fil85}, or Seyfert 2s (e.g., Spinelli et
al. 2006).  
The AGN nature of this object has been constantly debated, an aging
starburst being a compelling alternative \citep{maoz95, mao05, era02,
  koerding05}.  The galaxy presents a ring of H {\sc ii} regions at a
radius of $\sim 50$\arcsec, red arcs at $\sim 15$\arcsec, a high
surface brightness nuclear region, and high far-infrared bulge
emission \citep{kinney93, smith94}.  
Its intricate structure of off-nuclear compact source detections in
X-ray \citep{era02}, radio \citep{koerding05} and UV \citep{mao05},
that do not necessarily match with each other, certainly increase the
ambiguity associated with the nature of the main nuclear ionization
mechanism.  A common implied scenario in all of these studies is that
this system is probably in the final stages of a merger.

We reexamine here NGC 4736 in the context of additional evidence for
its AGN nature, which is the detection of a broad H$\alpha$ component
in its nuclear spectrum, as observed by the {\it Hubble Space Telescope} ($HST$) with the Space Telescope Imaging Spectrograph (STIS).  We also gather multi-wavelength data and show that source of the broad H-alpha emission line is coincident with a compact X-ray and radio source.   These observations suggest that NGC~4736 hosts a broad line region of significantly low luminosity, which makes this object one of the least luminous LINERs with strong evidence for BH accretion.  Interestingly, BH mass indicators calibrated on rapidly accreting Seyfert galaxies give highly discrepant mass estimates in this more quiescent system.  
The $\sim 10^7 M_{\sun}$ value given by the $M-\sigma^*$ is two orders of magnitude higher than the values obtained via estimators based on the observed emission (X-ray variability, scaling relations, mid-IR emission) which, although physically independent of each other, give a consistent result of $\sim 10^{5} M_{\sun}$.  
We are thus facing the following conundrum: either a) the standard AGN BH mass indicators do not necessarily apply to sources emitting in this low luminosity regime as, probably, the emission mechanism is fundamentally different from that associated with higher luminosity AGN, or b) the emission signatures do not trace accretion onto the central BH.  We discuss possible resolutions of this discrepancy in \S5, where we propose some rather exotic scenarios.



\section{Data compilation \& analysis} \label{data}

NGC 4736 has been quite extensively observed across
the whole electromagnetic spectrum.  We present in this section the
multiwavelength observations of this galaxy nucleus, in connection to new
measurements of the broad H$\alpha$ emission detected with $HST$-STIS.   

\subsection{The broad H$\alpha$ emission  observed with {\it HST}-STIS} \label{broad}  

High resolution optical spectra of the NGC 4736 nucleus were obtained
with $HST$-STIS on July 2002.\footnote{Data are publicly available (Prop
  ID 8591) but have not been published.}  The observations were
obtained with the $52\arcsec \times 0\farcs1$ aperture oriented at
PA$=49.65 ^{\circ}$, with the slit centered along the major axis of the
starlight distribution; two cosmic ray split exposures were obtained,
one being slightly shifted in the slit direction.  The total combined
exposure time is $\sim$4000s.  The G750M grating was set at 6581\AA,
with a scale of 0\farcs05/pixel, with no binning.  We reduced the
spectra using IRAF\footnote{IRAF is distributed by NOAO, which is
  operated by AURA Inc., under contract with the National Science
  Foundation} and the STIS reduction pipeline maintained by the Space
Telescope Science Institute \citep{dressel07}. This reduction included
image combination and cosmic-ray rejection, flux calibration and
correction of the wavelength to the heliocentric frame.
To measure the nuclear nebular line-emission properties we extracted
the 1-dimensional aperture spectrum five pixels wide ($0\farcs25$)
centered on the continuum peak.  The extracted spectrum thus consists
of the central emission convolved with the STIS spatial point-spread
function (PSF) and sampled over a 
rectangular aperture of $0\farcs25 \times 0\farcs1$.
The measurements span $6295 - 6867$\AA\ with a resolution of
0.87\AA\ ($\sigma_{\rm inst} = 17$ km s$^{-1}$).

\begin{figure}
\epsscale{1.2}
\plotone{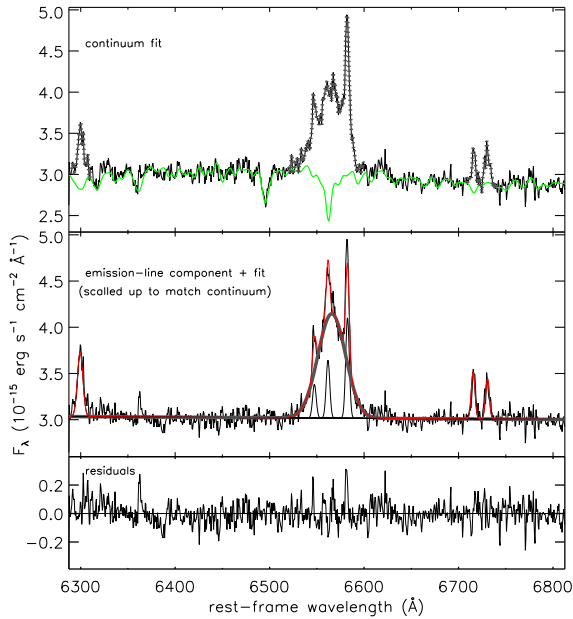}
\caption{Spectral fits to the HST-STIS spectrum of NGC 4736, based on a 5 pixel extraction aperture, which corresponds to a total emission area of 0.025 arcsec$^2$.  {\sl Upper panel}: The thin continuous line shows the observed data, while the green line shows the best fit stellar population model obtained by clipping out the hashed emission-line region.  
{\sl Middle panel:} The continuum subtracted spectrum, i.e., the emission-line component, together with the corresponding spectral fit, that have been shifted by an additive constant to match the initial continuum level, for illustration purposes.   The thin continuous lines show the
  individual Gaussian components (1 per narrow line), the thick grey
  continuous line represents the broad H$\alpha$ feature, and the red
  line is the final fit to the observed spectrum.  
  {\sl Lower panel: }The residuals after the subtraction of the model fit. 
\label{fitbroad}}
\end{figure}

Figure~\ref{fitbroad} shows the resulting spectrum together with the best matching continuum model, spectral fits of the pure emission-line component, and the associated residuals.    
The continuum model of the underlying stellar population is obtained via a $\chi^2$ minimization of a 
 nonnegative least-squares fit between the observed spectrum and a sum of discrete star bursts of different ages, adopted from the \citet{bru03} stellar population synthesis templates, 
 together with dust attenuation modeled as an additional free parameter.   The continuum fitting is performed using an adaptation to our data of Christy Tremonti's code \citep{tremonti04}.
 The pure emission-line spectrum, obtained by subtracting the modeled continuum from the observed spectrum,  is fit by a combination 
 of linear continuum and Gaussian components.
In addition to narrow emission, the
H$\alpha$+[\ion{N}{2}] feature shows clear evidence for a broader
feature.   A flux ratio
of 1:3 was assumed for the [\ion{N}{2}] doublet, as dictated by the
branching ratio \citep{ost89}; the [\ion{O}{1}] feature and the [\ion{N}{2}] and
[\ion{S}{2}] doublets were assumed to share a common velocity
(red)shifts and widths.  
 The best-fitting Gaussian parameters were
derived via an interactive $\chi^2$ minimization, using SPECFIT
\citep{kriss94}.  Because of the generally low signal to noise of the 2-d spectrum we are not able to test whether our measurements of the broad H$\alpha$ feature could be corrupted by a possible rotating disk as in the case of M84 (e.g., Walsh, Barth, \& Sarzi 2010); the match in the widths of the [S II] and [N II] narrow features (as long as that of the narrow H$\alpha$) argue however against such a significant effect.

\begin{deluxetable}{lcl}
\tablecolumns{3} 
\tablewidth{0pt} 
\tablecaption{Broad H$\alpha$ Emission Measurements
\label{tbl-broad}} 
\tablehead{ 
\colhead{quantity} &
\colhead{value} & 
\colhead{notes}}
$f_{\rm blend}$      &0.80&fraction of broad H$\alpha$ to H$\alpha$+[\ion{N}{2}] blend. \\
$f_{\rm H\alpha}$    &0.93&fraction of broad H$\alpha$ to total H$\alpha$ \\
FWHM(H$\alpha^{\rm broad}$)& 1570&  $\pm$ 110; in km s$^{-1}$ \\
$\Delta$v            & 140 & $\pm$ 20; in km s$^{-1}$; broad relative to narrow H$\alpha$\\
log $F(\rm H\alpha^{\rm broad})$&-13.38& observed flux (erg s$^{-1}$ cm$^{-2}$)\\
log $L(\rm H\alpha^{\rm broad})$& 37.96& observed luminosity (erg s$^{-1}$)\tablenotemark{a}
\enddata 
\tablenotetext{a} {assuming distance $d = 4.3$ Mpc.}
\end{deluxetable}

Measurements related to the broad H$\alpha$ feature from the STIS
measurements are listed in Table ~\ref{tbl-broad}.  The fractional
contribution of this broad component to the total flux of the
H$\alpha$+[\ion{N}{2}] blend is 80\%.  Compared to other galaxies with
broad line emission in the Palomar survey \citep{ho97IV}, the width
of the broad H$\alpha$ is typical (1570~km s$^{-1}$).  However, at
just 9.1$\times10^{37}$~ergs~s$^{-1}$, the H$\alpha$ luminosity is
lower than any known broad line sources in the Palomar sample or the
Sloan Digital Sky Survey \citep{ho97IV,greene07}.  This includes the famous low
luminosity Seyfert 1 galaxy, NGC~4395, which has a luminosity of
$1.2\times10^{38}$~ergs~s$^{-1}$ \citep{filippenko89} and is at a
similar distance to NGC~4736.
Note also that the broad H$\alpha$ feature is redshifted by
$\sim 140 \pm 20$ km s$^{-1}$ relative to the narrow emission, that is
locked at the systemic velocity of the object; velocity offsets
between the broad H$\alpha$ and narrow lines are not unusual
\citep{ho97a}.   

Evidence for broad H$\alpha$ emission has also been presented in a recent
PCA tomography study applied to this nucleus \citep{steiner09}.  The
observations reported in this case come from the Gemini Multi Object
Spectropraph (GMOS)-IFU data cube, and have been obtained 4 years
after the $HST$-STIS spectrum had been acquired.  Thus, the broad
H$\alpha$ line associated with the nucleus of M94 appears to be
persistent for at least 4 years.  They derive a broad H$\alpha$
luminosity of $6\times10^{38}$~ergs~s$^{-1}$, $\ga$ 6 times brighter
in the GMOS data than the one we detect in the STIS spectrum.  The
difference could result from true variability in the broad line
luminosity, or could be due to differences in aperture size or
placement.  We note that \citet{steiner09} suggest the broad line
region is offset from the photocenter of the galaxy by 0\farcs15,
while our spectral $0\farcs25 \times 0\farcs1$ aperture coincides with
the photocenter.  Even with their higher H$\alpha$ luminosity, this
source is among the least luminous known broad line AGN.


\subsection{Ground-based optical spectroscopy \& the spectral classification}

Ground-based optical spectra of the NGC 4736 nucleus are available from the Palomar spectroscopic survey by \citet{ho97a} and the integrated spectrophotometric survey survey of \citet{mous06} conducted with the 2.3m Bok telescope, as well as  from new data we acquired on February 2008 at the MMT Observatory.  The MMT spectrum  is obtained with the Blue Chanel Spectrograph, the 500 grooves/mm grating used in first order with the 1\arcsec slit, and covers $\lambda\lambda3800 - 7000$ with 3.6\AA\ resolution.  None of these data show indications of broad H$\alpha$ emission.  This outcome is not surprising given the feature's flux in the {\it HST} spectrum, which would be very difficult to discern in the 1\arcsec or 2\arcsec apertures employed in these ground-based observations, which are at least one order of magnitude larger than that used in the $HST$-STIS observations.    Through a 2\arcsec aperture ($\sim$ 40 pc), NGC 4736's emission complex H$\alpha +$ [\ion{N}{2}] is generally heavily swamped by the host stellar light.


\begin{figure}
\epsscale{1.3}
\plotone{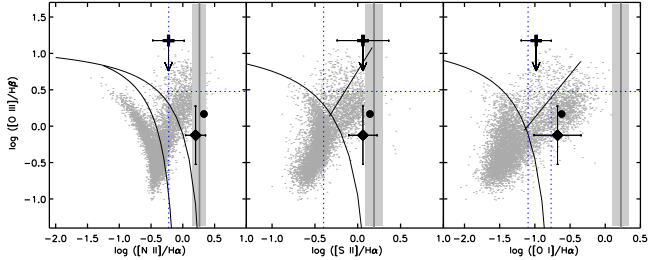}
\caption{NGC 4736 within emission-line diagnostic diagrams.  Filled circles
 reflect measurements based on the Palomar observations.  Diamonds represent measurements from
 \citet{mous06}, while crosses show measurements from our MMT
 spectrum.   The vertical bands indicate the HST-STIS measurements and the associated errors (with no correction for reddening; [\ion{O}{3}] and H$\beta$ are not
 available in the HST spectra).   The solid (black) curves indicate the \citet{kew06} 
 classification, while the dotted (blue) lines indicate criteria used by
 \citet{ho97a}.   The background grey points correspond to measurements of SDSS nearby galaxies from \citet{con08}. 
\label{bpt}}
\end{figure}

Probably as expected, measurements of NGC 4736's narrow-line emission are consistent with this system being powered at least partially by a non-stellar source, regardless of the resolution of its observation.  Figure
~\ref{bpt} shows the location of this nucleus in the 3-dimensional diagnostic
diagram usually employed in classifying emission-line galaxies
\citep{bpt,vei87,kew06}, for different sets of data mapping $\sim10$ to $\sim100$ pc.  
As with the $HST$-STIS observations, the ground-based measurements are performed after the host stellar
contribution has been subtracted from the observed spectrum by means of absorption galaxy
template fits.   Our MMT spectrum offers only an upper limit for
the [\ion{O}{3}]/H$\beta$ while the Palomar line flux measurements
are at least 50\% uncertain.  We list the measurements of the nebular
emission of this nucleus as observed by Palomar, Bok, MMTO, and
$HST$-STIS in Table ~\ref{tbl-fl}.   Because measurements of the [\ion{O}{3}]/H$\beta$ ratio are not available in the {\it HST} spectrum, we show the high-resolution measurements in Figure ~\ref{bpt} using the [\ion{O}{3}]/H$\beta$ from the Palomar catalog; it is readily apparent that  the high resolution data are consistent with a LINER or a Seyfert classification for this object, and reveal its type 1 (broad-lined) character.

\begin{deluxetable}{lcccc}
\tablecolumns{5} 
\tablewidth{0pt} 
\tablecaption{Emission Line Measurements\tablenotemark{a}
\label{tbl-fl}} 
\tablehead{ 
\colhead{line name} &
\colhead{HST-STIS} & 
\colhead{Palomar\tablenotemark{b}} &
\colhead{Bok\tablenotemark{c}} &
\colhead{MMTO\tablenotemark{d}}}
\startdata
~[\ion{O}{1}] $\lambda$6300&  5.03$\pm$0.31  &  6.03&  2.8$\pm$0.9 & 7.0$\pm$1.0\\
~H$\alpha$ (narrow)              &  3.00$\pm$0.32  & $>25.12$& 13.5$\pm$1.9 & 22.0$\pm$2.4\\
 ~[\ion{N}{2}] $\lambda$6583&  5.25$\pm$0.25  & 54.01& 21.5$\pm$1.5 & 39.3$\pm$2.9\\
~[\ion{S}{2}] $\lambda$6716&  2.48$\pm$0.18  & 18.59&  8.3$\pm$1.1 & 15.5$\pm$0.3\\
~[\ion{S}{2}] $\lambda$6731&  2.16$\pm$0.18  & 16.33&  7.2$\pm$1.1 & 9.8$\pm$0.3
\enddata 
\tablenotetext{a} {All fluxes are in units of 
$10^{-15}$ erg s$^{-1}$ cm$^{-2}$, and represent the observed 
values, not corrected for reddening.}
\tablenotetext{b}{nonphotometric conditions; line ratios are at
  least 50\% uncertain.}
\tablenotetext{c}{2$\farcs$5 slit, 2.3m telescope; photometric conditions \citep{mous06}.}
\tablenotetext{d} {1\arcsec slit; fluxes are in units of $10^{-15}$ erg s$^{-1}$ cm$^{-2}$.}
\end{deluxetable}

\subsection{The radio, UV and X-ray observations \& the astrometric coincidence with the broad H$\alpha$ emission}

 Across the electromagnetic spectrum, the center of this galaxy exhibits a complex morphology.   There is a plethora of bright   radio, UV and X-ray sources detected in the center of NGC 4736.  There is a nuclear compact (15 GHz) radio source measured by \citet{nag02} that appears to be associated with that of the brightest of the two close (8.49 GHz) radio sources revealed by ~\citet{koerding05}, and to that of the brightest of the two ultraviolet point sources, which varies on a 10-year timescale \citep{mao05}.  {\it Chandra} observations \citep{era02} reveal
numerous discrete X-ray sources in the inner galaxy; the second
brightest X-ray source, X2 ($L_{X,2-10 keV} = 5.9 \times 10^{38}$ erg
s$^{-1}$), coincides within the errors with the nucleus position.  The off-nuclear radio, UV and X-ray sources are apparently unrelated to each other. 

The relation to possible optical counterparts of these observations is
not well constrained in the literature.  We thus investigated this
issue, with a particular interest in the degree to which our newly
detected broad H$\alpha$ coincides in position with the
multi-$\lambda$ detections.  We used for this purpose a variety of
archival images from the {\it HST}: WFPC2/PC data in the F555W filter
data (PID: 5741 and 10402), HRC data in the F250W and F330W filters
(PID: 9454), and NICMOS/NIC3 data in the F160W filter (PID: 9360).
All data were downloaded from the {\it HST} archive and images were drizzled
together when required.  Absolute astrometry was performed on these
data as well as the STIS observations (taken with the slit out)
using a ground based V-band image obtained from the {\it Spitzer
  Infrared Nearby Galaxies Survey} (SINGS) ancillary data
\citep{kennicutt03}.  These SINGS observations were aligned to the USNO-B system
using $\sim$80 stars and then astrometry of the {\it HST} data was obtained
by degrading the resolution of the F555W image to match the SINGS
image.  All other {\it HST} images were then matched with the
astrometry corrected F555W frame to an accuracy of $<0\farcs05$.  The
absolute error on this astrometry is about 0$\farcs$2, and is dominated by
scatter of stars in the SINGS image relative to the USNO-B positions.


\begin{figure*}
\begin{center}
\leavevmode
\epsfxsize=15cm
\plotone{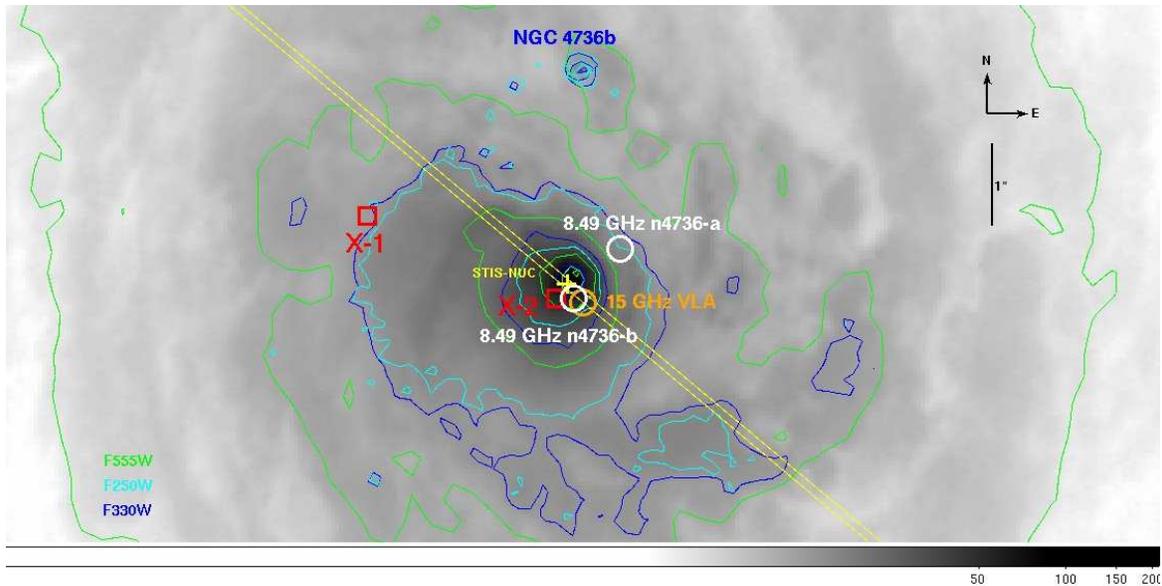}
\caption{ The mosaic-ed F555W (V-band; combination of 5 WFPC2 images
  taken at two different epochs) image of the NGC4736 nucleus, with
  intensity contours overlaid (green).  The position of the
  STIS-nucleus is indicated (by the yellow slit) along with the locations of the 2 most
  luminous and closest to the nucleus hard X-ray sources, X1 and X2, 
  as detected with Chandra (red squares), along with
  the position of the 15  GHz VLA radio core 
  \citep[orange circle]{nag02}, and those of the two radio compact
  sources detected by \citet{koerding05} (white circles).  The blue and cyan
  contours correspond to the F330W and F250W observations from
  \citet{mao05}; note the presence of NGC 4736b, $\approx 2\farcs5$
  to the north of the nucleus.  The absolute 
astrometry is good to within 0$\farcs$2 (pixel size is
0$\farcs$05 for F555W, and 0$\farcs$025 for F250W and F330W). 
\label{m94nuclei}}
\end{center}
\end{figure*}

Figure ~\ref{m94nuclei} illustrates the result of this data
compilation and the corresponding radio, UV and X-ray source matches.
{\em It is clear that, within $< 0\farcs2$ (4~pc), there is an obvious
astrometric match in the nuclear X-ray, UV, optical, radio compact
sources, and the newly detected broad H$\alpha$ emission line.} The
nucleus position as observed by STIS has RA: 12$^h$50$^m$53$\fs$20,
DEC: 41$\degr$ 07$\arcmin$ 13$\farcs$40.
The off-nuclear X1, that is the brightest compact X-ray source detected in this galaxy nucleus, has no counterpart at other wavelengths. Same is true for the off-nuclear 8.49 GHz detection, that is only 1\arcsec ($\sim$20pc) away from the nucleus. The off-nuclear UV source is the only one detected in optical light.

\section{NGC 4736's nuclear emission across the electromagnetic spectrum and $L_{\rm bol}$} \label{sed_txt}

\begin{figure}
\epsscale{1.2}
\plotone{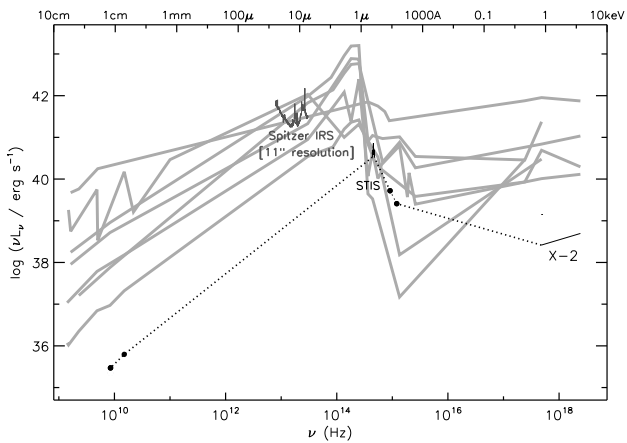}
\caption{Radio to X-ray nuclear SED of NGC 4736 superposed on SEDs of LLAGN from
  \citet{ho99}.  The optical STIS spectrum stands out as an amorphous blob featuring the strong H$\alpha$ emission feature.  
  The 2-10 keV Chandra X-ray detection X2 is depicted as a power-law corresponding to $\Gamma = 1.6$.  The low spatial resolution Spitzer IRS observations are shown for comparison.
  \label{sed_all}}
\end{figure}

With the wealth of data available for this galaxy nucleus, we are able
to build its least contaminated X-ray to radio nuclear spectral energy
distribution (SED).  
The multiwavelength observations of the sources detected in the very central regions of NGC
4736, plotted in $\nu L\nu$ units, are displayed in Figure
~\ref{sed_all}.  The  X-ray detection X2 is represented here by a
power-law, estimated based on its photon-indices $\Gamma=1.6$, where the
absorbing column density is fixed at the Galactic value.  Both F250W
and F330W UV observations are plotted, only for the nuclear detections.  
The optical data are represented by the
{\it HST} -STIS spectrum, featuring the strong H$\alpha$ emission line.
Measurements corresponding to all the radio observations discussed above are
indicated.  For the sake of completeness, we also include in this plot lower spatial resolution
observations from SINGS  IRS
\citep{kennicutt03} as well. We show all of these measurements superposed
on the SEDs of LLAGN from \citet{ho99}.  No artificial normalization
has been performed, and there is no correction for absorption in
either the Galactic extinction (except for the X-ray data), or
intrinsic to NGC 4736.


\begin{deluxetable}{rcccc}
\tablecolumns{5} 
\tablewidth{0pt} 
\tablecaption{Nuclear SED Data
\label{tbl-sed}} 
\tablehead{ 
\colhead{log ($\nu$/Hz)} &
\colhead{log ($\nu L_{\nu}$/erg/s)} & 
\colhead{Resolution} &
\colhead{Obs. date} &
\colhead{Instr.}} 
\startdata
18.38 & 38.69& 0$\farcs$15& 5/2000& Chandra-ACIS\\
17.68 & 38.42& 0$\farcs$15& 5/2000& Chandra-ACIS\\
15.08 & 39.41& 0$\farcs$5& 6/2003& ACS F250W\\
14.96 & 39.72& 0$\farcs$5& 6/2003& ACS F330W\\
14.68 & 39.91& 0$\farcs$1& 6/2002&HST-STIS\\
14.64 & 39.93& 0$\farcs$1& 6/2002&HST-STIS\\
13.48 &40.58 & 4$\farcs$7& 5/2004&IRS-SH\\ 
13.19 &40.33 & 11$\farcs$1& 5/2004&IRS-SH/LH\\ 
12.91 &40.87 & 11$\farcs$1& 5/2004&IRS-LH\\ 
10.18 & 35.79& 0$\farcs15$& 1/2001& VLA, A config. \\
 9.93 & 35.47& 0$\farcs24$& 6-10/2003& VLA, A config.
\enddata 
\end{deluxetable}

With X2 and n4736-b as the X-ray and the 8.5 GHz counterparts of the
broad H$\alpha$ detection respectively, and assuming the continuum
could be described as simple power-laws between the points we present
data for, this object's nuclear SED corresponds to $L_{\rm bol}
\approx 2.5 \times10^{40}$ erg s$^{-1}$.  Spitzer data is not included
in estimating $L_{\rm bol}$ because they do not reflect the IR
emission of the uncontaminated nucleus; the aperture used in these
observations is very large (see Table~\ref{tbl-sed}), including the
entire field-of-view shown in Figure \ref{m94nuclei}. 
This luminosity makes NGC 4736 one of the least luminous LINERs with strong
evidence of BH accretion.  This source is thus
a critical signpost of BH accretion at extremely low levels.
Note that this object's weak emission is most likely not caused by obscuration;
\citet{ho97a} list a Balmer decrement of $H\alpha/H\beta = 3.1$ (albeit highly uncertain, with a probable error of $\pm100\%$),  and \citet{era02} provide an upper limit for the neutral
hydrogen column density of $N_H < 3.3 \times 10^{20}$ cm$^{-2}$.

\subsection{Comparison with other observed LLAGN}

Figure~4 shows that NGC~4736's SED is very similar to those of other LLAGN \citep{ho99}.  This suggests that, in spite of the significant difference in $L_{\rm bol}$, there is no fundamental transition in the accretion mode in this source compared to other LLAGN.

The similarity of the SED means that previously proposed bolometric
corrections for LLAGN appear to work quite well for NGC~4736.  The
value estimated based on the correction to the [\ion{O}{3}] line
luminosity that \citet{hec04} proposed to work well for Seyferts is
$L_{\rm bol} \approx 3500 \times L_{\rm [O III]} = 4.5 \times 10^{40}$
erg s$^{-1}$ ( here $L_{\rm [O III]}$ is measured from the ground
within the Palomar survey, and is not corrected for intrinsic
reddening).  If the more recent assessment of the $L_{\rm bol} = 600
\times L_{\rm [O III]}$ bolometric correction of \citet{kauffmann09}
is used, with $L_{\rm [O III]}$ corrected for reddening based on the
Balmer decrement listed above and a $\tau \propto \lambda^{-0.7}$
attenuation law \citep{charlot00}, then $L_{\rm bol} = 0.93 \times
10^{40}$ erg s$^{-1}$, which compares satisfactorily with the measured
$L_{\rm bol}$.  The average bolometric correction to the observed 2-10
keV X-ray luminosity proposed by \citet{ho08}, with $L/L_X = 16$,
results in a  somewhat lower value, $0.94 \times 10^{40}$ erg s$^{-1}$; with the more recent $L/L_X = 50$ average bolometric correction to the 2-10 keV luminosity of \citet{eracleous10a}, $L_{\rm bol} = 1.8 \times 10^{40}$ erg s$^{-1}$, and thus, very close to our integrated value.
Within a typical uncertainty of $\sim$50\%, all of these estimates are consistent with the measured $L_{\rm bol}$ value.   
Other, more uncertain bolometric indicators are also consistent with our integration: the correlation between the mid-IR [\ion{Ne}{5}] emission-line luminosity and bolometric luminosity derived by \citet{satyapal07} for a small sample of  much brighter nearby AGN,  that has a large scatter ($\sim$ 1 order of magnitude), gives a  $L_{\rm bol} = 1.6 \times 10^{40}$ erg s$^{-1}$, which is very close to our measured value.

This good match among quite a variety of bolometric estimators derived independently from the multiwavelength properties of AGN supports the AGN interpretation for this system.  However, there are two details of its emission spectrum which are unlike typical broad-lined LLAGN.

First, despite that fact that the broad H$\alpha$ emission of NGC~4736 is one of the weakest measured among type 1 AGN, the total H$\alpha$ emission is peculiarly strong relative to the X-ray counterpart. 
The ratio $L_X$(2-10 keV)/$L_{\rm H\alpha}$ is $\sim$6, and thus {\it lower} than the median of $\sim 15$ exhibited by the type 1 AGN (and low-$z$ quasars) included in ~\citet{ho01} study that revealed a relatively tight correlation between the two types of emission.   With the higher broad H$\alpha$ flux from \citet{steiner09}, the ratio is even lower, $L_X$(2-10 keV)/$L_{\rm H\alpha} \approx 1$.       
This finding is surprising given that the low luminosity nearby galaxies that deviate from the $L_X - L_{\rm H\alpha}$  linear scaling toward lower values of the $L_X/L_{\rm H\alpha}$ ratios are the type 2 sources, mostly Transition Objects, where the ionization mechanism is not necessarily dominated by an AGN-type of source.
Following \citet{ho01} arguments, the unusually low $L_X/L_{\rm H\alpha}$ ratio measured in the nucleus of NGC 4736 suggests that either: 1) the optical line emission is not powered exclusively by a central AGN, or 2) the X-ray emission in this system arises, at least partially, from a non-AGN source, e.g., an X-ray binary.  A final possibility is that the source is highly variable (as suggested by the UV, X-ray, and H$\alpha$ observations) and thus, the unusual $L_X/L_{\rm H\alpha}$ is the result of measurements made at different times.

Second, NGC 4736's nebular emission shows a number of peculiarities. 
To start with, the electron density in NGC~4736's nebular emission
is very low, and it does not show the typical gradient exhibited by AGN, i.e., increase toward the more nuclear regions (Constantin et al. 2011, in prep.).  In both the Palomar and {\it HST} -STIS observations, the ratio [\ion{S}{2}] $\lambda\lambda$ 6716/6731 is $\sim 1.26$ (Table ~\ref{tbl-fl}), implying an electron density of $n_e \sim 10^2$ cm$^{-3}$, which lies at the lowest end in the distribution of particle densities measured in the Palomar objects \citep{ho03}.   

Given the lack of a density gradient in the line emitting region it is then rather peculiar to observe a significant increase (by a factor of $\sim$6) in the [\ion{O}{1}]/H$\alpha$ line flux ratio in the {\it HST}  spectrum relative to the large aperture observations, which would usually be interpreted as an indication for a more pronounced AGN-like ionization in the more central regions.   There are two reasons for this: 1) [\ion{O}{1}]  requires a significantly hard radiation field, i.e., that of an AGN, to sustain a sufficiently extensive partially ionized zone in clouds optically thick to Lyman continuum, and thus to produce a strong such feature. Since the ionization potential of [\ion{O}{1}] matches that of H very well, large differences in the [\ion{O}{1}] /H$\alpha$ ratios are expected between accretion and non-accretion sources.
2) given the lack of a density stratification in this nucleus,  the degree of ionization of the emitting gas is expected to diminish with radius as the ionizing radiation emerging from a nuclear source falls off in density as $r^{-2}$, and thus produce  strong gradients in the [\ion{O}{1}] /H$\alpha$ ratio.

In summary, NGC 4736 may be a broad-lined AGN, with an SED similar to other LLAGN, but it is atypical in several aspects whose physical origin remains unclear.   Given the above listed possible explanations for the peculiar $L_X/L_{\rm H\alpha}$ ratio, we evaluate in the next subsection whether photoionization by the weak AGN in this system is sufficiently powerful to balance the emission  cooling in this system.


\subsection{Comparison of ionizing and emission-line power}


The multi-wavelength observations of this system allow for a relatively rigorous assessment of whether the photoionization by this system's AGN can power the measured emission-line luminosities, and in particular that of the broad H$\alpha$ component.  
Following \citet{eracleous10}, we can run an energy budget test via a direct comparison of the H$\alpha$ luminosity and count rate with the ionizing luminosity $L_i = L_{\rm 1Ry -- 100 keV}$ and the ionizing photon rate $Q_i = Q_{\rm 1Ry -- 100 keV}$.    

It is important to treat separately the broad and narrow emission line features as there is strong evidence that they originate from regions of significantly different physical conditions.  The broad H$\alpha$ comes from a much more compact and much denser emitting gas than the narrow Balmer and forbidden lines.   The difference in density is at least 3 orders of magnitude; the critical density for collisional excitation of the [\ion{O}{1}]  ($\sim 2 \times 10^6$ cm$^{-3}$) can be used to estimate the gas density of the broad line region, as [\ion{O}{1}] does not exhibit a broad component in this nucleus' emission.  The difference in size is expected to be $\sim 2$ orders of magnitude (e.g., reverberation mapping of nearby Seyfert galaxies measured to be $<1$ week, Denney et al. 2010).  Thus, the mechanisms that can operate in these two regions are expected to be qualitatively different.  The photon and energy balance conditions should reflect these differences, and thus should differ as well; in particular, potential contributions to the H$\alpha$ emission via collisional excitation should be minimal for the narrow line emitting region, but important for the broad component.

Based on photoionization models (i.e., Cloudy, v94.0; see Ferland et al. 1998) computed by \citet{lewis03} for a wide range of ionization parameters, densities and metallicities, \citet{eracleous10} finds that energy balance in a line-emitting nebula requires that $L_i > 18 (\pm 2) L_{\rm H\alpha}/f_c$, where $f_c$ is the covering factor, or the fraction  of the ionizing luminosity of the AGN that is absorbed by the line-emitting gas.    These models are covering electron densities that appear to encompass both the broad and the narrow line emitting regions in NGC 4736;  thus,  with only  a fraction of 10\% of ionizing photons photons being absorbed by the line-emitting gas, the minimum energy balance condition for AGN ionization is given by $L_i/L_{\rm H\alpha} > 180$, for both narrow and broad emission features.     
In the same time however, a minimum requirement for photon balance can be quite different for the two emission regions: for the narrow line component $Q_i > 2.2~ Q_{\rm H\alpha}$, corresponding to the case B recombination (i.e., one H$\alpha$ photon is emitted for every 2.2 recombinations);  for the denser broad line emitting region, the number of H-alpha photons that can be produced for each ionization can be at least 7-8 times higher than the standard case B estimate, or  $Q_i \ga 0.25~ Q_{\rm H\alpha}$\citep[e.g.,][]{ost89}.



\begin{deluxetable}{lcccc}
\tablecolumns{5} 
\tablewidth{0pt} 
\tablecaption{Ionizing and Emission-Line Power
\label{tbl-ionization}} 
\tablehead{ 
\colhead{} &
\colhead{log $L_{\rm H\alpha}$\tablenotemark{a}} &
\colhead{log $Q_{\rm H\alpha}$\tablenotemark{a}} &
\colhead{$L_i/L_{\rm H\alpha}$\tablenotemark{b}} &
\colhead{$Q_i/Q_{\rm H\alpha}$\tablenotemark{c}} }
\startdata
Palomar 			& $>37.7$&$>49.3$& $<93$	& $<2.75$  \\
Bok 				& 37.5       & 49.0	& 165.9	& 4.9 \\
MMTO			& 37.7	& 49.2	& 104.7	& 3.1 \\
HST-STIS, narrow    & 36.8	& 48.4	& 759	&  22.4\\
HST-STIS, broad      & 37.9	& 49.5	& 57.5 	& 1.6 \\
PCA tomography      & 38.8	& 50.3	& 8.7		& 0.25 
\enddata 
\tablenotetext{a}{the luminosities are measured in erg s$^{-1}$ and the photon rates in s$^{-1}$.}
\tablenotetext{b}{ the minimum energy balance condition for AGN ionization is given by
 $L_i/L_{\rm H\alpha}$  $\ga 180$ (when a fraction of 10\% of ionizing photons photons is absorbed by the line-emitting gas)}
\tablenotetext{c}{the minimum photon balance condition for AGN ionization is given by $Q_i/Q_{\rm H\alpha} > 2.2$ for the narrow line regions, and $Q_i/Q_{\rm H\alpha} \ga 0.25$ for the broad component. }
\end{deluxetable}

With the $L_i$ and  $Q_i$ values already calculated by \citet{eracleous10} by integrating M94's nuclear SED assuming that pairs of points could be connected by a power-law, and with the H$\alpha$ measurements from all {\it HST} and ground-based observations presented above, we can proceed with the comparison.    Table ~\ref{tbl-ionization}   lists the  $L_{\rm H\alpha}$, $Q_{\rm H\alpha}$, and associated ratios $L_i/L_{\rm H\alpha}$, $Q_i/Q_{\rm H\alpha}$ for all of these optical spectroscopic observations, where the narrow and broad H$\alpha$ measurements are shown separately.

It is readily apparent that while $L_i/L_{\rm H\alpha}$ is in general $>18$ ($f_c=1$), it is almost never $>180$ ($f_c=0.1$); the only exception is the narrow H$\alpha$ emission measured in the  $HST$ aperture.   Thus, a dominant AGN ionization of the narrow line region in this nucleus is definitely possible in the $HST$ aperture, but only for $f_c \ga 18\%$ at larger radial distances.    For the $HST$ broad H$\alpha$ feature, an AGN ionization  is possible only if $f_c \ga 30\%$.  The PCA tomography measurement of the broad H$\alpha$ argues, however, against a balanced energy budget originating entirely in an AGN-like power mechanism, even when a maximum covering factor is considered; the AGN-produced energy falls short of the required amount by at least 50\%.    Interestingly, the measured $Q_i/Q_{\rm H\alpha}$ ratio is well within the required photon balance corresponding to an AGN excitation for both the narrow and the broad emission features. 
Thus, with one clear exception, the AGN in NGC 4736 appears to be capable of providing enough photons to explain the observed H$\alpha$  luminosity but only for relatively high $f_c$ values.

Simply because the AGN-like SED of NGC 4736 can explain the majority of the ionization energy and photon rate does not imply that  the actual mechanism is an AGN;  alternative excitation mechanisms must be explored.  Possible options are:  i) we are missing ionizing photons from accretion onto the central BH, or ii) there are other power sources that could make up the power deficit in this system, particularly for producing the broad H$\alpha$ feature.

The first alternative could be possible if we were observing an ``echo" of a previous epoch of more violent accretion, a few hundred years ago.   This idea has been explored by Eracleous et al. (1995), who showed that the reverberation of an ionizing flare in the nebula can produce LINER-like emission-line ratios.   In this scenario, it is expected that the central UV source and the  [\ion{O}{3}]  line would also follow the decay of the ionizing continuum; while the UV observations are not providing clear evidence for such a decay over the course of one decade, the multiple optical spectroscopic observations that we present in Figure ~\ref{bpt} are consistent with a possible decrease in time (a few years) in the [\ion{O}{3}] flux (in this model, H$\beta$, [\ion{S}{2}], [\ion{O}{1}] are expected to decay very slowly, in $60-250$ years).   Nevertheless,  this duty-cycle hypothesis also requires that the broad-line emission 
fades immediately if the ionizing continuum declines, and the PCA tomography observations show that this is not happening in this source.   The echo of such an ionizing continuum flare should also be detectable in a narrow-band [\ion{O}{3}] $\lambda$5007 image in the form of a ring, which is not readily observed in M94. 

The second alternative of power sources other than accretion, has been often proposed in explaining the emission-line spectra of LINERs, with mixed success (see \S1).    The most probable alternative sources appear to be the mechanical power delivered by compact radio jets, along with photoionization by young or post-AGB stars from old or intermediate-age stellar populations.      Shock models are highly unlikely to produce broad emission features, and thus, are not favored in this case.   
Recent star-formation activity remains however a viable options, particularly in light of relatively new discoveries of peculiar supernovae with broad H$\alpha$ features that do not appear to fade in time.  This idea is discussed in more detail in \S5.

\section{The Black Hole Mass} \label{bhmass}


Given the unusual energetics of M94's nucleus, it is important to investigate whether the BH mass estimators derived for rapidly accreting Seyferts, which appear to be widely used for AGN, also work in the low-luminosity regime flagged by this particular system. 

There are a variety of indirect methods that can be used to estimate
the mass of BHs in galaxy centers.   The available multi-wavelength measurements of the nuclear emission for  NGC~4736 allow the calculation of $M_{\rm BH}$ based on four different techniques, as well as with the $M - \sigma^*$ relation.     In this section, we explore and compare the results of these five methods, along with their consequences for this object's energetics, and a comparison with a recent dynamical measurement of $6.68 (5.14 - 8.22) \times 10^6 M_{\sun}$ for this object, which is listed in \citet{kormendy11} as obtained from Gebhardt et al. (2011, in preparation).   We first present the methods and associated $M_{\rm BH}$ calculations, and then discuss the shortcomings of each measurement.

\begin{enumerate}
\item Using the $M - \sigma^*$ relation, established for quiescent
  nearby galaxies, including both ellipticals and spirals with
  classical bulges, as quantified by \citet{gultekin09}, for $\sigma^*
  = 110 \pm 5$ km s$^{-1}$ \citep{bar02}, $M_{\rm BH} = 1.05 \pm 0.64
  \times 10^7 M_{\sun}$; within errors, this value agrees well with Gebhardt et al. dynamical measurement.  With this value, $L_{\rm bol}/L_{\rm Edd} \approx 2 \times 10^{-5}$, which is consistent with the range of values within LINERs are expected to lie \citep{ho04,ho08}.    

\item Using the {\it HST}  measurements of the FWHM and the luminosity of the broad H$\alpha$
  component (see Table ~\ref{tbl-broad}) within the scaling relation based
  solely on observations of this broad emission feature, as derived by
  \citet{gre05}, we obtain $M_{\rm BH} \approx 3 \times 10^4 M_{\sun}$;
  the fractional uncertainty associated with this measurement is
  $\sim$30\%, and includes both the scatter in the scaling relation
  and the errors in the line measurements.  With the broad H$\alpha$
  line luminosity measured in the PCA tomography study
  \citep{steiner09}, the BH mass would increase (by a factor of 6.5)
  to $M_{\rm BH}=1.9 \times 10^5 M_{\sun}$.  The difference in these
  two values could be considered the most conservative uncertainty
  associated with this BH mass estimate.  
The corresponding Eddington ratios for these BH masses are $L_{\rm
  bol}/L_{\rm Edd} \approx 1 - 7 \times 10^{-3}$.

\item Mid IR detection and measurements of the [\ion{Ne}{5}] (14.32
  $\mu$m) and [\ion{O}{4}] (25.89 $\mu$m) emission lines
  ~\citep{dud09} give, via the empirical correlations between the MIR
  line luminosities and reverberation mapping-based $M_{\rm BH}$
  values presented by \citet{dasyra08}, a black hole mass of $2.3 \pm
  0.4 \times 10^5 M_{\sun}$, and $1.7 \pm 0.5 \times 10^5 M_{\sun}$
  respectively.  
  The corresponding Eddington ratio is in this case
  $L_{\rm bol}/L_{\rm Edd} \approx 1.0 \times 10^{-3}$.  Note that the
  scatter adopted for these relations is only a lower limit of the
  real value, thus, the uncertainty may be larger.

\item The normalized X-ray excess variance method, as described in
  \citet{papadakis08}, applied to Chandra observations \citep{era02}
  gives $M_{\rm BH} = 2.5 \pm 1.7 \times 10^5 M_{\sun}$.  To be
  specific, we used for this calculation the excess variance
  $\sigma=0.06 \pm 0.04$, $L_{\rm bol} = 2.5 \times 10^{40}$ erg
  s$^{-1}$ as derived in Section \ref{sed_txt}, and  $\nu_{\rm lf} =
  1/T = 1/14$ h$^{-1}$ (where $T$ is the length of the light curve), to estimate the break frequency $\nu_{\rm bf}$ and then $M_{\rm
    BH}$ via equations 4 and 6 respectively, of \citet{papadakis08}.

\item The $M_{\rm BH}$ of this system can also be obtained via the ``fundamental plane of black hole activity" that relates black hole mass to the emitted compact radio $L_R = \nu L_{\nu}$(5GHz) and hard X-ray luminosities $L_X$(2-10 keV), and spans nine orders of magnitude in black hole mass.  We calculate $L_R =  1.7 \times 10^{35}$ erg s$^{-1}$ using the two 8.5GHz and 15 GHz radio measurements presented in this paper (Figure \ref{m94nuclei}, Table \ref{tbl-sed}), with a flux modeled by a power-law $S_{\nu} \propto \nu^{-\alpha}$.   Using the empirical fits of \citet{merloni03} for the fundamental plane relation, log ($M_{\rm BH}/M_{\sun}) = 5.9 \pm 1.1$ for the black hole in this system;  note that this measurement and its associated errors embrace all of the estimates presented above, and thus, do not provide any additional constraint to the $M_{\rm BH}$.  Interestingly, the latest derivation of the fundamental plane relation \citep {gultekin09b}, applied to very low nuclear galactic luminosities, provides a log ($M_{\rm BH}/M_{\sun}) = 7.2 \pm 0.4$ for this system, which remains consistent only with the $M - \sigma^*$ value, as it departs considerably from  those provided by the relations employing AGN emission. 

\end{enumerate}

There is a significant inconsistency between the value given by the $M - \sigma^*$ relation, which is supported by Gebhardt's dynamical measurement, and those based on the AGN emission.   
The $M - \sigma^*$ relation suggests a BH mass in NGC~4736 of 1$\times$10$^7$~M$_\sun$, and a correspondingly low $L_{\rm bol}/L_{\rm Edd}$ of $\sim2\times10^{-5}$.  Three of the estimates based on the AGN emission converge to black hole masses of $\sim10^5$~M$_\sun$ (and $L_{\rm bol}/L_{\rm Edd} \sim 10^{-3}$), showing a surprisingly consistent departure of two orders of magnitude from the $M - \sigma^*$ estimate.   The fundamental plane relation provides a $M_{\rm BH}$ value right in between these two different situations,  however, with no real additional constraint, due to its associated large uncertainty.    
These differences are somewhat puzzling given that the BH mass estimates based on nuclear emission properties are all calibrated to follow the $M - \sigma^*$ for high mass black holes ($M_{\rm BH} \ga 10^6 M_{\sun}$).  It is however true that the calibrators are biased toward nearby Seyfert galaxies with much higher Eddington ratios than that of NGC 4736.
We briefly discuss in the following subsections more specific weaknesses of each measurement.

\subsection{Caveats of the  $M - \sigma^*$ relation} \label{mbhsigma}
The $M - \sigma^*$ relation is expected to provide a reliable estimate
of the BH mass as it is based on the strong correlation between
dynamical mass measurements of supermassive BHs and their host
properties \citep{ferrarese00, gebhardt00, tre02,gultekin09}.  This
relationship is derived primarily from ellipticals and spirals with
classical bulges (formed during major mergers).  Recent observations
suggest it may not be valid for samples of later-type spirals which
more commonly host pseudobulges (formed via secular disk processes)
\citep[e.g.][]{greene10}.  Because the distinction between classical
and pseudobulges is based on formation, it does not simply correlate
with observable properties \citep{kormendy04,fisher10}.  In general
pseudobulges are less luminous, have lower bulge-to-total ratios, have
on-going star formation and lower S\'ersic indices than classical
bulges.   The possibility that the NGC~4736 is in fact a
pseudobulge provides a solution to the apparent conflict in BH
mass estimates.

Nevertheless, pseudobulges are difficult to identify, and there is not yet a consensus as to what defines them.  To complicate things further, classical and pseudobulges can exist within the same galaxy (e.g., NGC 2787; Erwin et al. 2003).
The classification of the NGC 4736 bulge is ambiguous:
\citet{fisher10} classifies it as a pseudobulge based on its nuclear
spiral and bar and low S\'ersic index ($n = 1.3$).  However, they also
find it has a low star formation rate, more typical of classical bulges,
and thus classify it as an ``Inactive Pseudobulge;''  if NGC4736 hosts a pseudobulge, it is an atypical one. 
We have created a surface brightness profile from NICMOS and 2MASS H-band data from
the Large Galaxy Atlas \citep{jarrett03} and found results that
conflict with the fits of \citet{fisher10}.  Specifically, we find
S\'ersic indices for the bulge of $n = 2.3 - 3.0$ depending on the
radial range and type of fit (single vs. double S\'ersic), which are consistent with a classical
one \citep[as shown in][]{fisher10}.  

Our surface brightness profile fits also reveal the presence of a
nuclear star cluster within the central $\sim 0.6\arcsec$ ($\sim$12pc), with 
an H-band magnitude of $\sim$12.5.  Such nuclear star clusters are
common in early-type spiral galaxies \citep{carollo02}.  The
luminosity and mass of nuclear star clusters are known to scale with
bulge luminosity and mass \citep{balcells03,ferrarese06,rossa06}, and
the NGC 4736 nuclear star cluster has a luminosity that is 0.1\% of
its bulge, typical for nuclear star clusters \citep{cote06}.  Nuclear
star clusters commonly co-exist with black holes, but there are a very
limited number of cases where masses for both can be estimated
\citep{seth08}.  In these cases, including the Milky Way, the BH mass
is similar to the mass of the nuclear star cluster within an order of
magnitude.  There is also some evidence that the ratio of BH mass to
nuclear star cluster mass increases with spheroid mass
\citep{graham09}.  For NGC 4736, assuming an old population with an
H-band M/L $\sim$ 0.7, the nuclear cluster would have a mass of
$\sim2\times10^7$~M$_\odot$.  This mass is quite similar to the $M -
\sigma^*$ BH mass estimate, suggesting thus the presence of a similar
sized BH.

\subsection{Caveats of the BLR scaling relation} \label{mbhhb}
The BLR scaling relations were derived using high luminosity
systems (i.e., Seyferts, not LINERs) and have been scaled to match the dynamical black
hole detections of BHs with masses $>10^6$~M$_\sun$.  Thus, their
applicability to NGC~4736, where they yield an estimate of
$\sim10^5$~M$_\sun$, is a (perhaps unwise) extrapolation.
These relations assume that the BLR is virialized due to proximity to
the BH \citep{peterson99} and thus the 5100\AA\ continuum luminosity
correlates with the emissivity weighted radius of the BLR, and thus with the BH mass
\citep{kaspi00}.  The overall errors associated with BH mass
measurements based on these relations do not exceed 0.5 dex
\citep{vestergaard02, nelson04, onken04}, and there appears to be good
consistency with BH masses obtained via the $M - \sigma^*$ relation
for relatively bright AGN ($L > 10^{42}$ erg s$^{-1}$), with $M_{\rm
  BH} \ga 10^6 M_{\sun}$ \citep{bar05}.  More recent calibrations
of the radius-luminosity relationship on which these techniques are
based infer that BH masses have been overestimated, however, only by
up to a factor of $\sim$3 \citep{bentz09}.  This latter study also
indicates a trend toward larger uncertainties, and larger amount of
overestimation in the BH with decreasing luminosity, however, no
conclusive results are yet available for this regime.

\subsection{Caveats of the MIR line-correlation} \label{mbhir}

The \citet{dasyra08} empirical relation between the MIR line emission
properties and the BH mass is derived using reverberation mapping BH
masses, and thus, as with the previous method, may not apply to
systems with $M_{\rm BH} < 10^6 M_{\sun}$.  
The relation holds for systems with $L_{\rm bol}/L_{\rm Edd} > 0.003$,
but not necessarily beyond this range.  The $M_{\rm BH} \approx 10^5
M_{\sun}$ given by this method places this object at the low end of
this $L_{\rm bol}/L_{\rm Edd}$ range; on the other hand, a more massive and
thus quiescent BH would correspond to a [\ion{Ne}{5}] luminosity of a
few orders of magnitude lower than that measured.  The scatter
associated with this relation remains in average 0.5 dex, and thus
cannot account for the $\sim 2-3$ orders of magnitude difference
between this value and that obtained via $M - \sigma^*$.

\subsection{Issues with estimating the BH mass from X-rays} \label{mbhx}
The BH mass derivation based on X-ray variability, or more precisely
on the relation between the excess variance and $M_{\rm BH}$, relies
on the hypothesis of a universal power spectral density function (PSD)
shape and amplitude in AGN, which is based
on the idea that the X-ray variability mechanism and the accretion
efficiency are the same for all AGN, at all redshifts.  
These assumptions appear to hold for the objects involved in deriving
and testing this relation \citep{papadakis08}; that sample is,
however, small and rather biased toward luminous (X-ray) sources, with
log $L_X/({\rm erg s^{-1}}) > 41.5$.  None of the 2 - 10 keV sources
detected in the nucleus of NGC 4736 are in this luminosity range.
Nevertheless, Galactic BHs ($M_{\rm bh} < 10^2 M_{\sun}$) in their
hard states show variability properties that match well those of AGN,
both of these types of sources falling on the same projection of the
$T_B - M_{\rm bh} - L_{\rm bol}/L_{\rm Edd}$ plane, where $T_B =
1/\nu_{\rm bf}$, suggesting that an extrapolation to the intermediate
mass BH (or lower $L$) level is practicable.


\subsection{Problems with the Fundamental Plane} \label{fplane}

Finally, it appears that the main problem with the fundamental plane is that the relation is not sharpened enough to provide strong constraints on BH masses.   Because the relation spans nine orders of magnitude in BH mass, it is expected to equally apply to any value of $M_{\rm bh}$ in the range we are interested in.     It is also the case that a wide variety of BH accretion models (e.g., with efficient and inefficient flows for the X-ray emission, or associating X-ray flux with synchrotron emission  near the base of a jet)  are consistent with this relation  (e.g., Falcke \& Biermann 1995; Heinz \& Sunyaev 2003), suggesting that a large diversity of accretion modes or rates are accommodated.   

Nevertheless, the fundamental plane remains best constrained only for systems with $M_{\rm BH} \ga 10^6 M_{\sun}$, and for very low nuclear luminosities, i.e., with negligible or zero AGN contribution that allow a dynamical measurement of their BH mass \citep {gultekin09b}.   In this regime, the latest derivation of the fundamental plane relation provides for M94 a $M_{\rm BH}$ value consistent with the $M - \sigma^*$ estimate and the dynamical measurement, while it would not reliably constrain a $M_{\rm BH} \la 10^6 M_{\sun}$ value.   
The scatter in the fundamental plane relation increases for lower BH masses, with higher Eddington ratios \citep {kording06, gultekin09b}. 



\section{Discussion: Alternative Power Generation Mechanisms} \label{discussion}

The detection of broad $H\alpha$ emission, combined with the spatial
coincidence of this emission with the detection of X-ray, UV, and
radio compact sources, provides strong evidence that this galaxy hosts
an accreting massive black hole.   The SED of this nucleus makes this object one of the lowest luminosity LINER with a distinct contribution to the total emission by black hole accretion.  In this scenario, the
nucleus is an AGN and the presence of off-nuclear sources,
particularly the radio and the UV detections, may result from remnant
jet activity emerging from the nucleus.
It is definitely exciting to detect AGN activity at energy levels
equal to that of several young supernova remnants of the Cas A variety
\citep{tur94}, an OB association which hosts a high mass X-ray binary
\citep{era02}, or simply a group of five late O supergiants
\citep{maoz95}.

Nevertheless, the presence of a number of unusual off-nuclear sources, coupled with the apparent deficit in the photoionizing photons, and with the fact that the BH mass estimates based on AGN emission appear to fail for this object, encourage us to explore alternative scenarios for the NGC~4736 nucleus.  
In particular, it is very likely that the BH mass estimates that exploit the multi-wavelength AGN characteristics don't work for this system because at least some of this emission is not the result of BH accretion.


There are certain (peculiar) kinds of core-collapse SNe that present multi-wavelength observations, and in particular broad H$\alpha$ emission components with characteristics that are very
similar to those we measure in the nucleus of M94.  The SN 2005ip,
presented by \citet{smith09}, is a very good example.  The so called
intermediate H$\alpha$ component associated with this SN ejecta
presents the same FWHM and brightness level as the broad H$\alpha$
detected in the STIS aperture, and it does not show any sign of
diminishing its strength over more than 3 year period (see their
Figure 7; a $very$ broad component with FWHM $\ga$ 10000 km s$^{-1}$ like the one exhibited by SN 2005ip
would not be measurable in the galaxy spectrum as it would be
completely swamped in the continuum stellar light, even in the
{\it HST}-STIS observations).

SNe as luminous as X2, with intrinsically
hard X-ray spectra (photon index $\Gamma \la 1$), have certainly been
encountered, for cases observed few years after the explosion (e.g.,
ATe \#1023; Pooley et al. 2007); for these cases, $L_X$/$L_{\rm bol}$ ratios are high relative to more standard SN cases \citep{immler07}.   
Another particular example of a SN that matches well the measurements of the nuclear emission in NGC 4736 is SDSS J09529.56+214313.3 \citep{komossa09}, which is believed (but
not confirmed) to be a SN type IIn.    For this system, the broad H$\alpha$ component stays strong for at least three years, and  its $L_X$(2-10 keV)/$L_{\rm H\alpha} \approx 1$, matching thus very well the surprisingly low value measured for NGC 4736.

The peculiar nebular, radio and UV characteristics of the nucleus of NGC 4736
appear to also compare well the SN phenomenology.   The electron densities (or
[\ion{S}{2}] line flux ratios) measured in this galaxy center are matching
exactly the ones measured in the environments of the extraordinary
type IIn SNe we compare here with, e.g., SDSS J09529.56+214313.3.  
The two different band radio observations of this
nucleus are consistent with emission from extragalactic SNe, in both
intensity and decline rate in flux density \citep{vandyk93,
  williams02, chandra09}.   Moreover, the compact radio off-nuclear detection,
only 1\arcsec (20 pc) away from the nuclear one \citep{koerding05}
could be interpreted as the result of shock emission associated with a nuclear
core-collapse SN.  Measurements of the brightness of the off-nuclear
UV detection in F250W and F330W place this object
  into the O star of late (5-ish) type spectral category, fitting thus well into the idea that this nucleus could simply be a star-forming site, and that NGC 4736a's emission includes significant contributions from a SN which exploded close to the weakly active (and massive, $M_{\rm BH}\sim 10^{(6-7)} M_{\sun}$) central BH.


This SN contribution scenario may have its own drawbacks.  If
the broad H$\alpha$ emission has actually gotten brighter by a factor of
6, as suggested by the comparison between the {\it HST}-STIS and GMOS
observations, the SN interpretation becomes problematic; variability of the broad line region originating in the AGN could, in principle, account for this effect.  Also, the
observed intra-day (hour-scale) X-ray variability measured for X2
remains yet to be detected in a supernova, and conflicts with the
physical scale over which X-ray emission is expected in SN remnants; thus, this particular behavior may remain strictly associated with the AGN.
Additional observations would be necessary to fully confirm or rule
out the SN scenario.
Specifically, new high resolution observations in the optical, UV, or
X-rays would be able to confirm if there is a fading in the light curve, as
expected from a SN, and would also allow accurate localization of
the source.  High S/N optical spectra would much better resolve
the emission-line profile, and a possible temporal evolution.

\section{Conclusion}

We have presented here an exhaustive multiwavelength analysis of the
nuclear emission properties of NGC 4736, prompted by new measurements of
a broad H$\alpha$ emission component detected in its high resolution {\it HST}-STIS
optical spectrum.  This broad H$\alpha$ component, with a luminosity
of $9\times10^{37}$ ergs s$^{-1}$, is one of the lowest luminosity broad line known.  
This broad H$\alpha$ is coincident with a compact bright X-ray
and radio source.  Our measurements of this object's 
spectral energy distribution reveal a bolometric luminosity of $L_{\rm
  bol} \approx 2.5 \times 10^{40}$ erg s$^{-1}$, that categorizes NGC
4736 as one of the least luminous LINERs with strong evidence for BH
accretion.   
Our comparison of five independent BH mass estimates reveals a discrepancy of two orders of
magnitude between the value $\sim 10^7 M_{\sun}$ predicted by the
$M - \sigma^*$ relation and the value $\sim 10^{5} M_{\sun}$ toward
which methods based on AGN emission activity in optical, mid-IR
and X-rays, seem to converge;  the fifth method is provided by the fundamental plane relation, which however, due to its large associated uncertainties, does not offer any additional constraint to this comparison.

We conclude that this system's BH mass cannot be reliably estimated via standard AGN BH mass indicators because the nuclear emission in this system is not entirely tracing the accretion onto the central BH.  Our assessment of the energy budgets of the ionizing and emission-line power suggests a possible deficit in the AGN ionization and production of a broad H$\alpha$ emission feature which can be made up by a peculiar kind of Type IIn SN that matches well the nuclear emission of NGC 4736 over the whole electromagnetic spectrum, and supports this galaxy nucleus' general aging starburst-like appearance.

\acknowledgements 

We thank the anonymous referee for constructive  comments that helped us improved the manuscript.
Support for this work was provided by NASA through grant number {\it HST}-AR-11749.01-A from the STScI, which is operated by the Association of Universities for Research in Astronomy, Inc., under NASA contract NAS5-26555. 

\clearpage

\end{document}